\documentclass[a4paper,preprint,superscriptaddress,preprintnumbers,amsmath,amssymb]{revtex4}

\usepackage[english]{babel}

\usepackage{graphicx}
\usepackage{epsfig}
\usepackage[dvips]{color}
\usepackage{float}

\usepackage{amsfonts,amsmath}

\usepackage{hyperref}

\newcommand{\dd}{\mbox{d}}

\begin{document}

\title{\bf{Pseudochaotic poloidal transport in the laminar regime of
the resistive ballooning instabilities}}

\author{I. Calvo}
\affiliation{Laboratorio Nacional de Fusi\'on, Asociaci\'on
EURATOM-CIEMAT, 28040 Madrid, Spain}

\author{L. Garcia}
\affiliation{Universidad Carlos III, 28911 Legan\'es, Madrid, Spain}

\author{B. A. Carreras}
\affiliation{BACV Solutions Inc., Oak Ridge, TN 37830,
U.S.A.}

\author{R. S\'anchez}
\affiliation{Fusion Energy Division, Oak Ridge National Laboratory,
Oak Ridge, TN 37831, U.S.A.}

\author{B. Ph. van Milligen}
\affiliation{Laboratorio Nacional de Fusi\'on, Asociaci\'on
EURATOM-CIEMAT, 28040 Madrid, Spain}

\begin{abstract}
In toroidal geometry, and prior to the establishment of a fully
developed turbulent state, the so-called topological instability of
the pressure-gradient-driven turbulence is observed. In this
intermediate state, a narrow spectral band of modes dominates the
dynamics, giving rise to the formation of iso-surfaces of electric
potential with a complicated topology.  Since ${\rm\bf E}\times{\rm\bf
B}$ advection of tracer particles takes place along these
iso-surfaces, their topological complexity affects the characteristic
features of radial and poloidal transport dramatically. In particular,
they both become strongly non-diffusive and non-Gaussian. Since radial
transport determines the system confinement properties and poloidal
transport controls the equilibration dynamics (on any magnetic
surface), the development of non-diffusive models in both directions
is thus of physical interest. In previous work, a fractional model to
describe radial transport was constructed by the authors. In this
contribution, recent results on periodic fractional models are
exploited for the construction of an effective model of poloidal
transport. Numerical computations using a three-dimensional reduced
magnetohydrodynamic set of equations are compared with analytical
solutions of the fractional periodic model. It is shown that the
aforementioned analytical solutions accurately describe poloidal
transport, which turns out to be superdiffusive with index $\alpha=1$.
\end{abstract}

\maketitle

\section{Introduction}

Numerical calculations of resistive pressure-gradient-driven
turbulence \cite{GarCarLyn} in toroidal geometry show the existence of
an unstable regime below the threshold for fully developed
turbulence. In this regime, the toroidal mode spectrum in steady state
is dominated by a single toroidal mode $N$. This dominant mode may
fluctuate intermittently among a narrow range of possible values. For
instance, with the parameters used in Ref.  \cite{GarCarLyn}, $24\leq
N \leq 26$. The transition from a stable plasma to this
ballooning-mode-dominated regime has the characteristic properties of
a topological instability
\cite{CarLynGarEdeZas,ZasCarLynGarEde}. After the transition, the
iso-surfaces of electrostatic potential induced by the ballooning
modes have a complicated topological structure, which is a direct
consequence of the (inner/outer) asymmetry in magnetic field strength
inherent to any magnetic toroidal geometry. Consequently, particles
advected by the ${\rm\bf E}\times{\rm\bf B}$ flows associated to these
complex iso-surfaces cease to behave diffusively. At this point, it is
worth stressing that this situation is quite different from cases of
near-critical turbulence discussed in the
literature~\cite{Carreras96}, in which the non-diffusive nature of
transport is related instead to the existence of spatio-temporal
correlations between fluctuations and the background profile gradients
from which they feed.

The structure of the potential iso-surfaces can be visualized as we
move in the toroidal direction around the torus, following the
magnetic field lines. At the singular magnetic surfaces, (potential)
vortices emerge with a structure which is consistent with the local
twist of the magnetic field lines. As the outermost part of torus (the
low-field side) is approached, filamentary vortices from different
singular surfaces may merge and form extended radial streamers. Radial
transport takes place predominantly within these streamers, since
particles can freely travel along them in the radial direction. The
nature of radial transport within these streamers characterizes the
confinement properties of the system. As we continue to follow the
lines toroidally, the vortices move back into the high-field side and
the streamers break up. Each particle will then remain trapped within
one of the filaments which emerge from this process. Poloidal (and
toroidal) transport then follows, as the population of particles
spreads out poloidally due to the free (ballistic) motion of each
particle along the radially localized filament which contains it. Its
nature thus determines how efficiently gradients are equilibrated on
any magnetic surface. Eventually, some of the filaments will reach
again the low-field side and merge to produce a new radial streamer,
enabling again radial transport until the structure goes back in to
the high-field side, brakes again up into new filaments, and ballistic
poloidal spreading ensues again. And so on.

In Ref.~\cite{ZasCarLynGarEde} the radial transport of tracer
particles in the unstable regime of this system was studied and found
to be of a fractional, self-similar nature. This means that their
transport can be described in terms of continuous-time random walks or
fractional differential equations with appropriate exponents. It was
also proven that the dynamics of tracers is pseudochaotic by showing
that the dispersion of trajectories of (initially close) particles is
polynomial (which in particular implies that the Lyapunov exponent is
zero). In addition, an analogy between the topological structure of
the flows and a billiard model (a paradigmatic example for
pseudochaos) was discussed at length.

In this paper we focus instead on the construction of an effective
transport model that captures the main features of poloidal
transport. This task is much more than a simple academic exercise or a
straightforward extension of the radial transport case. The reason is
that the difference with the radial case is not only the different
dominant physics, but the existence of a periodic boundary. In the
radial case, an absorbing boundary exists at the plasma edge, which
can be dealt with within the standard framework of fractional
differential operators. It has not been until very recently that the
mathematical framework necessary to deal with a periodic boundary has
been worked out by deriving the fluid limit equations of a
Continuous-Time Random Walk formulated on a circle
\cite{CalCarSanMil}. In the numerical calculations of
Ref. \cite{GarCarLyn}, the safety factor $q$ was taken to be between
$1$ (at the plasma center) and $2$ (at the plasma edge). Therefore,
for a dominant toroidal mode $N$, there are $N+1$ possible filaments
forming and the poloidal spreading of the tracers will take place in
up to $N$ finite-size steps in $q$.  As $N$ increases, the poloidal
spreading of particle tracers should approach the results for a
`fluid' transport model. In this paper, we will show that the new
framework can deal successfully with the problem of poloidal transport
in this system. It will also serve as illustration for future
applications regarding transport along periodic directions in any
magnetic configuration.

The rest of this paper is organized as follows. Section
\ref{sec:CTRWcircle} gives a survey of the main results of
\cite{CalCarSanMil} on Continuous Time Random Walks and the
fractional diffusion equation formulated on the circle. In Section
\ref{sec:numcalc} we present the reduced magnetohydrodynamic model
used in the study of the resistive pressure-gradient-driven
turbulence and compare the numerical and analytical results for
the transport of tracer particles. The conclusions are given in
Section \ref{sec:conclusions}. The appendix contains some basic
definitions on stable L\'evy distributions.

\section{Fractional diffusion equation on a circle}\label{sec:CTRWcircle}

Continuous Time Random Walks (CTRWs) \cite{MonWei,SchLax} are models
describing the microscopic transport of particles in a probabilistic
way. In this paper we are interested in the interpretation of poloidal
transport in fusion plasmas as a CTRW defined on a circle. A general
treatment of CTRWs on the circle has recently appeared
\cite{CalCarSanMil}. In this section we collect the results of
\cite{CalCarSanMil} which are relevant for the present work.

We denote by $n(\theta,t)$ the density of particles (tracer particles
for the application of the formalism relevant to this paper)
normalized to the total number of particles,
i.e. $\int_0^{2\pi}n(\theta,t)\dd\theta=1, \ \forall t$. The function
$n(\theta,t)$ must be periodic in $\theta$,
$n(\theta+2\pi,t)=n(\theta,t)$. A separable, Markovian, homogeneous
time-translational invariant CTRW is defined by a mean waiting time,
$\tau$, and a step-size pdf, $p(\Delta)$, giving the probability that
a particle performs a jump from $x$ to $x+\Delta$. The conservation of
probability requires that
$\int_{-\infty}^{\infty}p(\Delta)\dd\Delta=1$. Since $\Delta$ runs
over the interval $(-\infty,\infty)$, particles can wind around the
circle an arbitrary number of times in each jump. We are interested in
studying the case in which $p(\Delta)$ is a L\'evy stable distribution
(see Appendix \ref{sec:appLevy}). When the index of stability
$\alpha=2$, the pdf $p(\Delta)$ is Gaussian, whereas for $\alpha<2$ it
has algebraic tails. One would expect that (at least) in the latter
case the effect of the non-trivial topology of the circle be very
relevant and the CTRW on the circle should exhibit significant
differences with respect to a CTRW with the same step-size pdf
formulated on the real line.

The dynamics described by the CTRW defined above is equivalent to the
following Generalized Master Equation (GME) \cite{CalCarSanMil}:
\begin{equation}\label{eq:GMEper}
\partial_t n(\theta,t)= \frac{1}{\tau}\int_{0}^{2\pi}{\bar
p}(\theta-\theta')n(\theta',t')\dd \theta'-\frac{n(\theta,t)}{\tau},
\end{equation}
with
\begin{equation}\label{eq:defpbar}
\bar p(\theta) = \sum_{m=-\infty}^\infty p(\theta+2\pi m).
\end{equation}
Notably, $\bar p$ is obtained from $p$ by means of a ballooning
transform~\cite{Connor}. The sum in (\ref{eq:defpbar}) explicitly
accounts for the aforementioned fact that, given $\theta,\theta'\in
[0,2\pi)$, particles can arrive at $\theta$ from $\theta'$ through
jumps of length $|\theta-\theta'+2\pi m|$, $m\in{\mathbb Z}$.

Consider the case of a step-size pdf given by a stable L\'evy
distribution (see Appendix \ref{sec:appLevy}) with $\beta=0$, whose
characteristic function is:
\begin{equation}\label{eq:defLevyvar}
\hat p(k)=\exp( -\sigma^\alpha|k|^\alpha +i\mu k).
\end{equation}
Then, the fluid limit of
the GME (\ref{eq:GMEper}) is:
\begin{equation}\label{eq:GMEperFourierinhomflalphaneq3}
\partial_t n= -\frac{\sigma^\alpha}{2\tau\cos(\pi\alpha/2)}
({}_{0}{\cal D}^\alpha_\theta+{}^{2\pi}{\cal
D}^\alpha_\theta) n +\frac{\mu}{\tau}\partial_\theta n,
\end{equation}
where ${}_{0}{\cal D}^\alpha$ and ${}^{2\pi}{\cal D}^\alpha$ are the
    Riemann-Liouville operators on the circle derived in
    Ref.~\cite{CalCarSanMil}. The solution of
    Eq. (\ref{eq:GMEperFourierinhomflalphaneq3}) with initial
    condition $n(\theta,0)=\sum_{m=-\infty}^\infty\delta(\theta-2\pi
    m)$ (i.e. the propagator) is given by
\begin{equation}\label{eq:GMEperFourierinhomflhomogMarkLevy}
n(\theta,t)= \frac{1}{2\pi}\sum_{m=-\infty}^\infty
e^{(-\sigma^\alpha|m|^\alpha+i\mu m )t/\tau} e^{-im\theta}, \
\alpha\in(0,2].
\end{equation}
In the limit $t\to\infty$, $n(\theta,t)\to 1/2\pi$, as required by the
conservation of the number of particles.

When $\alpha=1$, which as we will see is the relevant case for the
present work, the infinite sum on the right-hand side of
Eq.~(\ref{eq:GMEperFourierinhomflhomogMarkLevy}) can be computed and a
closed expression for the propagator can be obtained:
\begin{equation}\label{eq:gensolalpha13}
n(\theta,t)=
\frac{1}{2\pi}\ \frac{\sinh(\sigma t/\tau)}{\cosh(\sigma
t/\tau)-\cos(\mu t/\tau-\theta)}.
\end{equation}

\section{Dynamical model and numerical calculations of tracer
particle transport}
\label{sec:numcalc}

In toroidal geometry, the underlying instability of the resistive
pressure-gradient-driven turbulence is the so-called resistive
ballooning mode \cite{Chaetal}.  To calculate the dynamical properties
of these instabilities, we use the reduced set of Magnetohydrodynamics
equations \cite{Str,DraAnt} in toroidal geometry, and because of the
low $\beta$ values, we also use the electrostatic approximation. The
model is then reduced to two equations: the perpendicular momentum
balance equation and the equation of state. The former can be written
in terms of the toroidal component of the vorticity $U$. In
dimensionless form it reads:
\begin{equation}\label{eq:U}
\frac{{\dd U}}{{\dd t}} = - S^2 {\bf{B}} \cdot \nabla \left( {\frac{{R^2
}}{{\eta F^2 }}{\bf{B}} \cdot \nabla \Phi } \right) + S^2 \frac{{\beta
_0 }}{{\varepsilon ^2 }}\frac{{{\bf{b}} \times {\boldsymbol{\kappa}}
}}{B} \cdot \nabla p + \mu \nabla _ \perp ^2 U.
\end{equation}
As for the equation of state:
\begin{equation}\label{eq:p}
\frac{{\dd p}}{{\dd t}} = D_{||} \frac{{R^2 }}{F}{\bf{B}} \cdot \nabla
\left( {\frac{{R^2 }}{F}{\bf{B}} \cdot \nabla p} \right) + D_ \perp
\nabla_\perp ^2 p.
\end{equation}
Here, $\dd /\dd t = \partial/\partial t + {\bf{V}}_\perp\cdot\nabla$ is the
convective derivative, and
\begin{equation}\label{eq:V}
{\bf{V}}_ \bot   =  - \frac{1}{B}\nabla \Phi  \times {\bf{b}},
\end{equation}
where $\Phi$ is the electrostatic potential.  There is a simple relation
between the toroidal component of the vorticity and the stream
function:
\begin{equation}\label{eq:relationUPhi}
U = \frac{1}{B}{\boldsymbol{\zeta}} \cdot \nabla  \times {\bf{V}}_ \bot
\end{equation}

Therefore, we solve Eqs.~(\ref{eq:U}) and (\ref{eq:p}) for $p$ and
$\Phi$, taking into account Eqs.~(\ref{eq:V}) and
(\ref{eq:relationUPhi}) to relate $U$ and $\Phi$. In Eq.~(\ref{eq:U}),
$\eta$ is the plasma resistivity, ${\bf b}={\bf B}/B$ is a unit vector
in the direction of the magnetic field, and
${\boldsymbol{\kappa}}={\bf b}\cdot\nabla{\bf b}$ is the magnetic
field line curvature.  The magnetic field is expressed as ${\bf
B}=F\nabla\zeta + \nabla\zeta\times\nabla\psi$, where $F = RB_\zeta$
is the toroidal flux function, which is a very slowly varying function
of the radial coordinate, $r$; $\psi$ is the poloidal flux, which is
not evolved in time, and $\zeta$ is the toroidal angle. Apart from the
dissipation terms, we have two dimensionless parameters in these
equations, $\beta_0= p(0)/(B_\zeta^2/2\mu_0)$ and
$S=\tau_R/\tau_{hp}$, the Lundquist number. Here, $\tau_R$ is the
resistive time at the magnetic axis, $\tau_R=\mu_0 a^2/\eta(0)$, and
$\tau_{hp}$ is the poloidal Alfv\'en time, $\tau_{hp}=R_0\sqrt{\mu_0
m_i n_0}/B_\zeta$, where $m_i$ is the ion mass, and $a$ and $R_0$ are
the minor and major radius, respectively. In the above dynamical
equations lengths are normalized to the minor radius $a$, and time to
the resistive time $\tau_R$.

To study the particle transport properties induced by these flow
structures, we use pseudo-particles as tracers.  These tracers are
solutions of the equation of motion:
\begin{equation}\label{eq:tracereq}
\frac{{\dd{\bf{r}}}}{{\dd t}} = {\bf{V}}_ \bot \left( {{\bf{r}},t} \right)
+ V_0 {\bf{b}}
\end{equation}
Here, the velocity is the flow velocity given by Eq.~(\ref{eq:V}) in
terms of the stream function, $V_0$ is an arbitrary velocity along the
field line, and ${\bf{b}} = {\boldsymbol{\zeta}} -
{\boldsymbol{\theta}}/q$. Since this model is electrostatic, all
information on turbulence evolution comes through the electrostatic
potential $\Phi$. In the present calculations we will use a constant
$V_0$ as initial condition for the tracer particles and keep the
velocity field frozen in time because we are only looking for the
effect of the flow structure on the transport.

The first results of particle tracer transport have been obtained
by launching $50000$ particles all with the same fixed initial
velocity $V_0 = 400 \pi$. At $t=0$ particles were located in a
small region around $r=0.7$, $\theta=0$, and $\zeta=0$. The
evolution is initially very asymmetric because of the preferential
direction induced by the drift in the motion of the particles
along the eddies, which are aligned with the field lines. However,
as particles go several times around the torus, $n(\theta,t)$
becomes increasingly symmetric. For this reason, we first remove
the drift of the distribution of tracers and then we compare the
evolution with the symmetric solution of the fractional diffusion
equation on the circle,
Eq.~(\ref{eq:GMEperFourierinhomflhomogMarkLevy}).

\begin{figure}[H]
\centering
\includegraphics[angle=0,width=10cm]{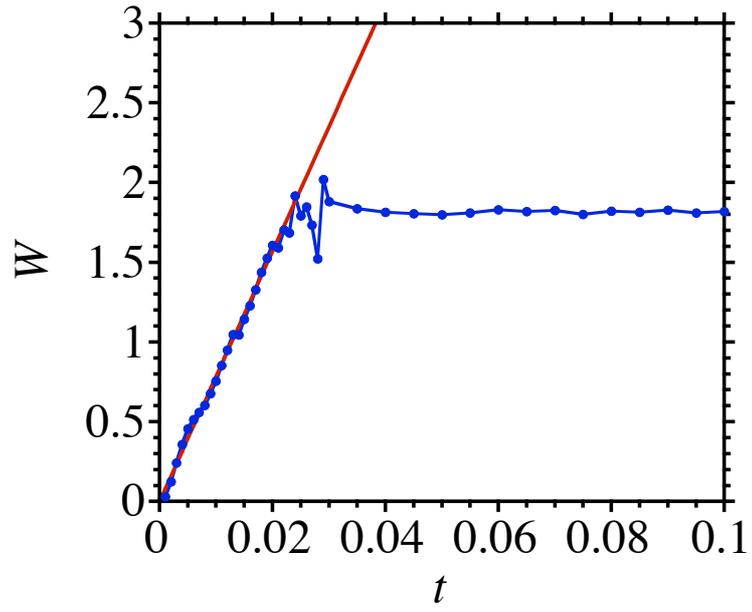}
\caption{Standard deviation of the pdf of particle tracers versus
time. The short-time behaviour implies that the index $\alpha$ is
equal to $1$.}
\label{FIG:1}
\end{figure}

\begin{figure}[H]
\centering
\includegraphics[angle=0,width=10cm]{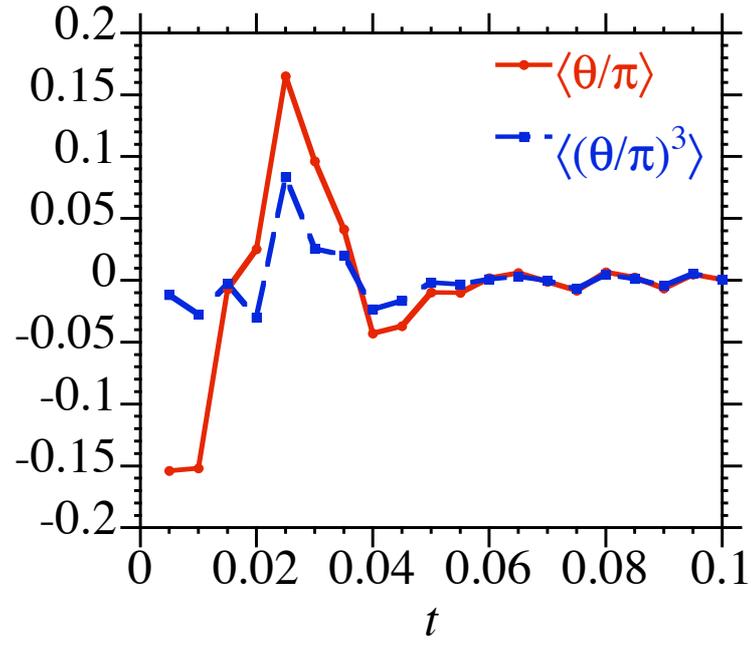}
\caption{First and third moments of the pdf of particle tracers
computed from the numerical integration of Eq.~(\ref{eq:tracereq}).}
\label{FIG:2}
\end{figure}

The Lyapunov exponent in the motion of the particle tracers is zero
and the trajectories separate from each other approximately linearly
in time. Therefore, we expect $\alpha$ to be close to $1$. Estimates
based on simplified models give $\alpha\in[1.05,1.11]$
\cite{ZasCarLynGarEde,ZasEde}. To test this assumption and to
determine the value of $\alpha$ associated with the poloidal
transport, we first look at the time evolution of the width of the
distribution, i.e. the square root of the second moment of the
particle distribution, $W(t)$. As shown in Fig.~\ref{FIG:1}, $W(t)$
increases linearly with time until it saturates at a constant value,
$W(\infty)=\pi/\sqrt{3}$, consistent with the flat tracer
distribution. This behavior was expected from the analytical
calculation. A fit of the data gives $\alpha= 0.984\pm 0.027$, so that
we can set $\alpha=1$. For the sake of completeness, we have plotted
in Fig.~\ref{FIG:2} the time evolution of the first and third moments
of the numerical pdf of tracers (after removing the drift), which are
essentially zero at any time.

\begin{figure}[H]
\centering
\includegraphics[angle=0,width=10cm]{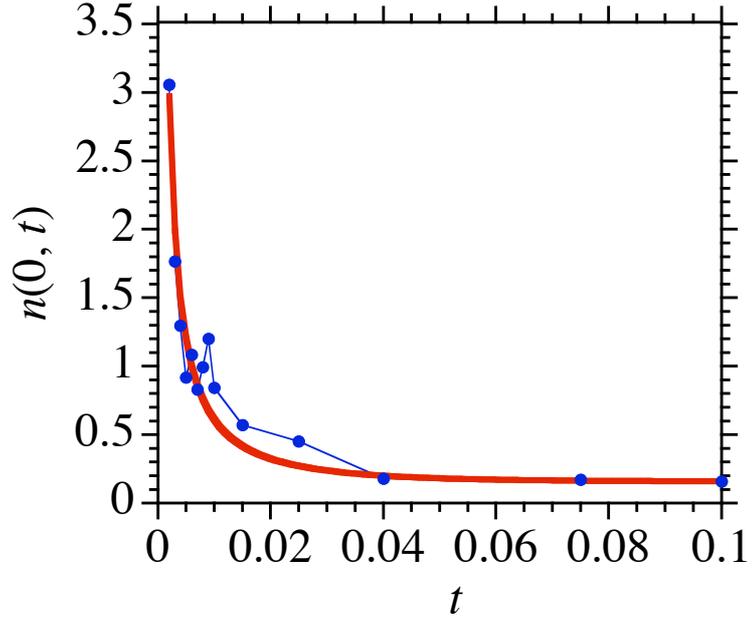}
\caption{Numerical results for $n(0,t)$ obtained from
Eq.~(\ref{eq:tracereq}) are fitted to the analytical expression
Eq.~(\ref{eq:nzero}), yielding $\sigma/\tau=0.268\pm0.013$.}
\label{FIG:3}
\end{figure}

\begin{figure}[H]
\centering
\includegraphics[angle=0,width=10cm]{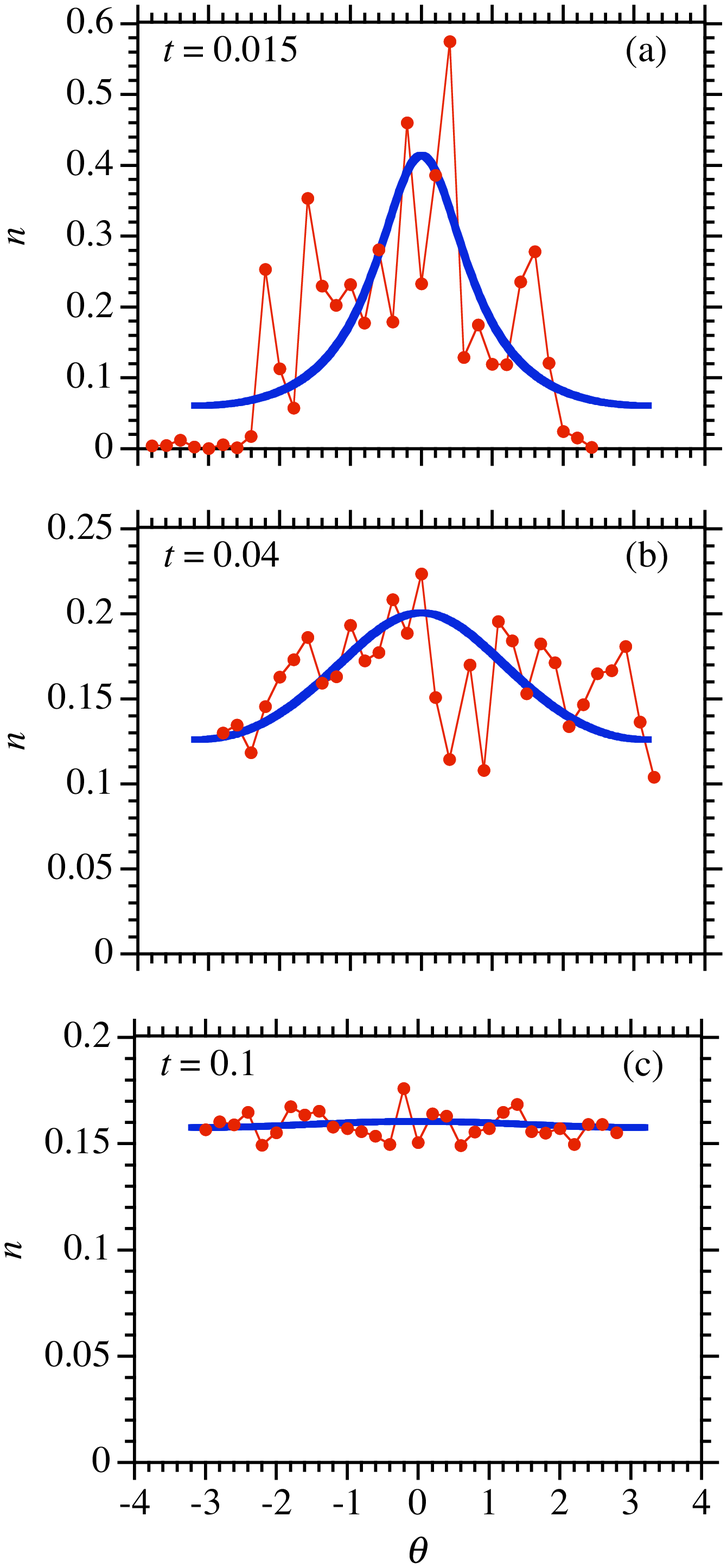}
\caption{Comparison of the time evolution of the pdf of tracer
particles with fixed initial toroidal velocity $V_0=400\pi$, and the
analytical expression, Eq.~(\ref{eq:gensolalpha13}).}
 \label{FIG:4}
\end{figure}

According to Eq.~(\ref{eq:gensolalpha13}), the expected time evolution
of the peak of the particle distribution is given by
\begin{equation}\label{eq:nzero}
n\left( {0,t} \right) = \frac{1}{{2\pi }}\frac{{1 + e^{ - \sigma
t/\tau } }}{{1 - e^{ - \sigma t/\tau } }}.
\end{equation}
By fitting this expression to the numerical data, we can determine the
time decay constant $\sigma/\tau$. The result of the fit is shown in
Fig.~\ref{FIG:3}, and the value of the constant is $\sigma/\tau =
0.268\pm 0.013$. Now, we can use the analytical expression for
$n(\theta,t)$ given by Eq.~(\ref{eq:gensolalpha13}) and compare this
analytical prediction with the numerical data.  We show a few examples
of this comparison in Fig.~\ref{FIG:4}. The analytical solution seems
to describe the numerical results relatively well. Recall that
Eq.~(\ref{eq:nzero}) is derived from a CTRW with symmetric step-size
pdf, hence the numerical distribution of tracers should have vanishing
odd moments, as is indeed the case (see Fig.~\ref{FIG:2}).

\begin{figure}[H]
\centering
\includegraphics[angle=0,width=10cm]{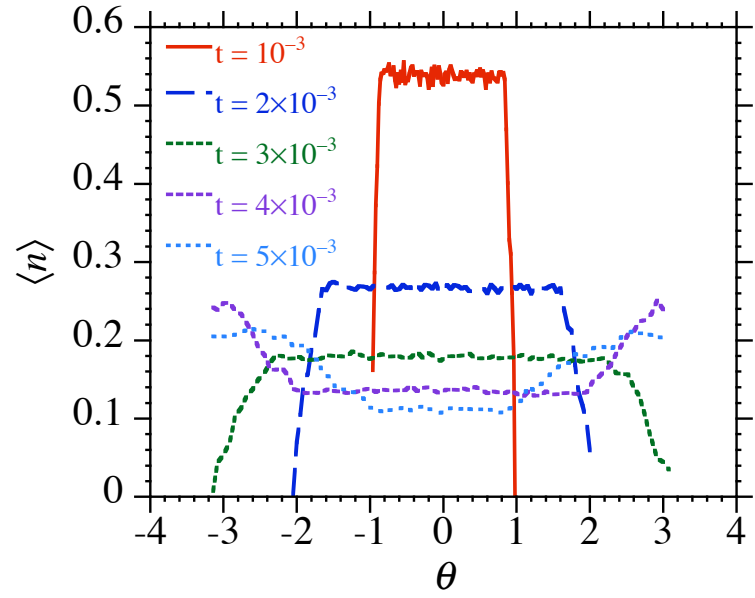}
\caption{Numerical computation of the time evolution of the pdf of
tracer particles with random initial toroidal velocities.}
\label{FIG:5}
\end{figure}

We have also calculated the time evolution of particle tracers with
random initial velocities in the toroidal direction. We have used
$50000$ particles and the results of a typical calculation showing the
evolution of $n(\theta,t)$ are plotted in Fig.~\ref{FIG:5}. In
Fig.~\ref{FIG:5}, one notes that the structures are gone. This is not
surprising and it is one of the expected consequences of the random
velocity initialization of the tracers. What may be surprising is the
change of the functional form of the distribution. These distributions
do not look at all like the analytical distributions given by Eq.
(\ref{eq:gensolalpha13}).

The explanation is simple. With a single initial velocity for all
particles, at any given time they were all located at the same
toroidal angle. Now, with the random initial velocities, at any given
time the particles are distributed over a range of toroidal angles and
what we have measured is a poloidal distribution averaged over all the
toroidal angles. Furthermore, since there is a mean drift associated
with the twist of the magnetic field lines, in each toroidal plane
$\zeta$, the poloidal distribution has its peak at a different value
of $\theta$. In each toroidal plane $\zeta$, the analytical
distribution is given by Eq. (\ref{eq:gensolalpha13}) with values a
given value of $\zeta$ in the range $[-V_{\rm{max}}t,
V_{\rm{max}}t]$. Here $V_{\rm{max}} = 400 \pi$ is the maximum velocity
of the tracers. Therefore, to reproduce the measured poloidal
distribution, we have to average the fixed $\zeta$ distribution over
the toroidal angle, that is
\begin{equation}
\langle n\rangle\left( {\theta ,t} \right) = \frac{{1 - e^{ - 2\sigma t/\tau }
}}{{2\pi }}\frac{1}{{2V_{\rm{max}} t}}\int\limits_{ - V_{\rm{max}}
t}^{V_{\rm{max}} t} {\frac{{d\zeta }}{{1 - 2e^{ - \sigma t/\tau } \cos
\left( {\theta - u\zeta } \right) + e^{ - 2\sigma t/\tau } }}},
\end{equation}
where $u$ is the average pitch of the field line at the radial
position of the initial tracers. After some algebra, one obtains
\begin{eqnarray}\label{eq:naveraged}
\langle n\rangle\left( {\theta ,t} \right) &= & \frac{1}{2\pi
uV_{\rm{max}} t}\Bigg\{ {\arctan\left[ {\frac{{1 + e^{ - \sigma t/\tau
} }}{{1 - e^{ - \sigma t/\tau } }}\tan \left( {\frac{{\theta +
uV_{\rm{max}} t}}{2}} \right)} \right]}\cr
&-&\arctan\left[ {\frac{{1 + e^{ - \sigma t/\tau } }}{{1 - e^{ -
\sigma t/\tau } }}\tan \left( {\frac{{\theta - uV_{\max } t}}{2}}
\right)} \right] \Bigg\}.
\end{eqnarray}

\begin{figure}[H]
\centering
\includegraphics[angle=0,width=10cm]{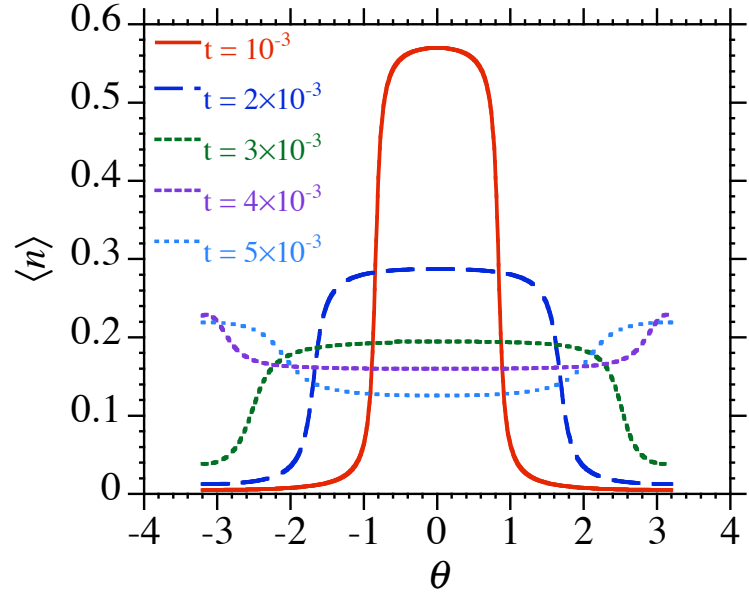}
\caption{Time evolution of $n(\theta,t)$ given by
Eq. (\ref{eq:naveraged}) for the values of the parameters
corresponding to the case plotted in Fig.~\ref{FIG:5}.}
 \label{FIG:6}
\end{figure}

This distribution should correspond to the one obtained in the
numerical calculations. Using the value of $\alpha$ obtained from the
previous section and the following estimate for $uV_{\rm{max}}$, $u =
2/3$ and $V_{\rm{max}} = 400 \pi$ as used in the numerical
calculations, we have plotted the corresponding pdfs in
Fig.~\ref{FIG:6}. The agreement between Figs.~\ref{FIG:5} and
\ref{FIG:6} is very good. For the randomized velocity case, once we
have averaged over the initial velocities, there is no difference
between anomalous diffusion with $\alpha=1$ and pure ballistic motion
of the particles in the absence of the flow structures. However, the
distinction is clear when we look at monoenergetic particles, as we
saw at the beginning of this section.

\section{Conclusions}
\label{sec:conclusions}

In this paper, we have used numerical simulations of resistive
ballooning mode turbulence at low beta (sufficiently so to produce a
topological instability) to illustrate the adequacy of recent periodic
formulations of fractional transport equations \cite{CalCarSanMil} to
describe non-diffusive turbulent transport in the poloidal
direction. The formalism is quite general and could be easily applied
to many other instances of transport problems along a periodic
direction, of which parallel equilibration dynamics in toroidal
magnetic devices is just one example.

In the case examined here, non-diffusive transport ensues due to the
complex topology of the potential iso-surfaces present in the system,
a direct consequence of the narrow spectral band of dominant
modes. Both radial and poloidal transport exhibit non-diffusive
features due to this complex topology, but their physical origin is
rather different, as explained in the text. The radial anomalous
diffusion in this system was already considered in
Ref.~\cite{GarCar06}. In the current paper, we have focused instead on
the poloidal transport, which turns out to be an interesting example
of $\alpha=1$ anomalous diffusion. The origin of this value of
$\alpha$ is in the fact that, away from the streamers, particles
simply spread out ballistically along the filaments while on the
high-field side. Once they reach the low-field side again and enter a
new streamer structure, their direction of motion can be modified as
the new streamer breaks up to form a new set of filaments towards the
high-field side and particles get trapped in them. It is this
combination of ballistic motion plus scattering of velocity directions
which is ultimately responsible of a Cauchy type (i.e., $\alpha =1$)
diffusive process.

By looking in detail at the poloidal distribution of monoenergetic
particle tracers, we have also shown that the new mathematical
framework can capture the features of the poloidal transport
process. There is very good agreement between numerical results and
the analytical calculation describing this phenomenon
\cite{CalCarSanMil}.  If tracers with random velocities are used, once
we average over initial conditions, it is not possible to distinguish
the anomalous diffusion with $\alpha=1$ from the normal ballistic
motion of the particles in the absence of the flow structures.

\begin{acknowledgments}
Research sponsored by DGICYT (Direcci\'on General de Investigaciones
Cient\'{\i}ficas y Tecnol\'ogicas) of Spain under Projects
No.~ENE2004-04319 and ENE2006-15244-C03-01 and by CM-UC3M (Comunidad
de Madrid - Universidad Carlos III) Project
No. CCG06-UC3M/ESP-0815. Part of this research was sponsored by the
Laboratory Research and Development Program of Oak Ridge National
Laboratory, managed by UT-Battelle, LLC, for the US Department of
Energy under contract number DE-AC05-00OR22725. L. G. acknowledges the
financial support of Secretar\'{\i}a de Estado de Universidades e
Investigaci\'on of Spain during his stay at Oak Ridge National
Laboratory.
\end{acknowledgments}

%%%%%%%%%%%%%%%%%%%%%%% APPENDICES %%%%%%%%%%%%%%%%%

\appendix

\section{L\'evy skew alpha-stable distributions}\label{sec:appLevy}

The family of {\it L\'evy skew alpha-stable distributions} (or simply
{\it stable distributions}, or {\it L\'evy distributions}) is
parameterized by four real numbers $\alpha\in(0,2]$, $\beta\in[-1,1]$,
$\sigma>0$, and $\mu\in{\mathbb R}$. Their characteristic function
(i.e. their Fourier transform) is given by \cite{Taqqu}:
\begin{equation}\label{eq:defLevy}
\hat S(\alpha,\beta,\sigma,\mu)(k)=
\left\{
\begin{array}{cc}
\exp\left(-\sigma^\alpha|k|^\alpha\left[1-i\beta\mbox{sign}(k)\tan(\frac{\pi\alpha}{2})\right]+i\mu k\right) & \alpha\neq 1,\\[8pt]
\hspace{-0.8cm}\exp\left(-\sigma|k|\left[1+i\beta\frac{2}{\pi}\mbox{sign}(k)\mbox{ln}|k|\right]+i\mu k\right) & \alpha = 1.
\end{array}
\right.
\end{equation}

According to the Generalized Central Limit Theorem \cite{Taqqu},
stable distributions are the only possible distributions with a domain
of attraction. The index $\alpha$ is related to the asymptotic
behaviour of $S(\alpha,\beta,\sigma,\mu)(x)$ at large $x$:
\begin{equation}
S(\alpha,\beta,\sigma,\mu)(x)=
\left\{
\begin{array}{cc}
C_\alpha\left(\frac{1-\beta}{2}\right)\sigma^\alpha|x|^{-1-\alpha} &
x\to -\infty,\\[8pt]
C_\alpha\left(\frac{1+\beta}{2}\right)\sigma^\alpha|x|^{-1-\alpha} &
x\to \infty,
\end{array}
\right.
\end{equation}
for $\alpha\in(0,2)$. For $\alpha=2$, $S(2,\beta,\sigma,\mu)$ is a
Gaussian distribution.

\newpage

\newpage

\newpage

\newpage

\newpage

\newpage

\begin{thebibliography}{14}
\expandafter\ifx\csname natexlab\endcsname\relax\def\natexlab#1{#1}\fi
\expandafter\ifx\csname bibnamefont\endcsname\relax
  \def\bibnamefont#1{#1}\fi
\expandafter\ifx\csname bibfnamefont\endcsname\relax
  \def\bibfnamefont#1{#1}\fi
\expandafter\ifx\csname citenamefont\endcsname\relax
  \def\citenamefont#1{#1}\fi
\expandafter\ifx\csname url\endcsname\relax
  \def\url#1{\texttt{#1}}\fi
\expandafter\ifx\csname urlprefix\endcsname\relax\def\urlprefix{URL }\fi
\providecommand{\bibinfo}[2]{#2}
\providecommand{\eprint}[2][]{\url{#2}}

\bibitem[{\citenamefont{Garcia et~al.}(2002)\citenamefont{Garcia, Carreras, and
  Lynch}}]{GarCarLyn}
\bibinfo{author}{\bibfnamefont{L.}~\bibnamefont{Garcia}},
  \bibinfo{author}{\bibfnamefont{B.~A.} \bibnamefont{Carreras}},
  \bibnamefont{and} \bibinfo{author}{\bibfnamefont{V.~E.} \bibnamefont{Lynch}},
  \bibinfo{journal}{Phys. Plasmas} \textbf{\bibinfo{volume}{9}},
  \bibinfo{pages}{47} (\bibinfo{year}{2002}).

\bibitem[{\citenamefont{Carreras et~al.}(2003)\citenamefont{Carreras, Lynch,
  Garcia, Edelman, and Zaslavsky}}]{CarLynGarEdeZas}
\bibinfo{author}{\bibfnamefont{B.~A.} \bibnamefont{Carreras}},
  \bibinfo{author}{\bibfnamefont{V.~E.} \bibnamefont{Lynch}},
  \bibinfo{author}{\bibfnamefont{L.}~\bibnamefont{Garcia}},
  \bibinfo{author}{\bibfnamefont{M.}~\bibnamefont{Edelman}}, \bibnamefont{and}
  \bibinfo{author}{\bibfnamefont{G.~M.} \bibnamefont{Zaslavsky}},
  \bibinfo{journal}{Chaos} \textbf{\bibinfo{volume}{13}}, \bibinfo{pages}{1175}
  (\bibinfo{year}{2003}).

\bibitem[{\citenamefont{Zaslavsky et~al.}(2005)\citenamefont{Zaslavsky,
  Carreras, Lynch, Garcia, and Edelman}}]{ZasCarLynGarEde}
\bibinfo{author}{\bibfnamefont{G.~M.} \bibnamefont{Zaslavsky}},
  \bibinfo{author}{\bibfnamefont{B.~A.} \bibnamefont{Carreras}},
  \bibinfo{author}{\bibfnamefont{V.~E.} \bibnamefont{Lynch}},
  \bibinfo{author}{\bibfnamefont{L.}~\bibnamefont{Garcia}}, \bibnamefont{and}
  \bibinfo{author}{\bibfnamefont{M.}~\bibnamefont{Edelman}},
  \bibinfo{journal}{Phys. Rev. E} \textbf{\bibinfo{volume}{72}},
  \bibinfo{pages}{026227} (\bibinfo{year}{2005}).

\bibitem[{\citenamefont{Carreras et~al.}(1996)\citenamefont{Carreras, Newman,
  Lynch, and Diamond}}]{Carreras96}
\bibinfo{author}{\bibfnamefont{B.~A.} \bibnamefont{Carreras}},
  \bibinfo{author}{\bibfnamefont{D.}~\bibnamefont{Newman}},
  \bibinfo{author}{\bibfnamefont{V.~E.} \bibnamefont{Lynch}}, \bibnamefont{and}
  \bibinfo{author}{\bibfnamefont{P.~H.} \bibnamefont{Diamond}},
  \bibinfo{journal}{Phys. Plasmas} \textbf{\bibinfo{volume}{3}},
  \bibinfo{pages}{2903} (\bibinfo{year}{1996}).

\bibitem[{\citenamefont{Calvo et~al.}(2007)\citenamefont{Calvo, Carreras,
  {S\'anchez}, and van Milligen}}]{CalCarSanMil}
\bibinfo{author}{\bibfnamefont{I.}~\bibnamefont{Calvo}},
  \bibinfo{author}{\bibfnamefont{B.~A.} \bibnamefont{Carreras}},
  \bibinfo{author}{\bibfnamefont{R.}~\bibnamefont{{S\'anchez}}},
  \bibnamefont{and} \bibinfo{author}{\bibfnamefont{B.~P.} \bibnamefont{van
  Milligen}}, \bibinfo{journal}{J. Phys. A: Math. Theor.}
  \textbf{\bibinfo{volume}{40}}, \bibinfo{pages}{13511} (\bibinfo{year}{2007}).

\bibitem[{\citenamefont{Montroll and Weiss}(1965)}]{MonWei}
\bibinfo{author}{\bibfnamefont{E.~W.} \bibnamefont{Montroll}} \bibnamefont{and}
  \bibinfo{author}{\bibfnamefont{G.}~\bibnamefont{Weiss}}, \bibinfo{journal}{J.
  Math. Phys.} \textbf{\bibinfo{volume}{6}}, \bibinfo{pages}{167}
  (\bibinfo{year}{1965}).

\bibitem[{\citenamefont{Scher and Lax}(1972)}]{SchLax}
\bibinfo{author}{\bibfnamefont{H.}~\bibnamefont{Scher}} \bibnamefont{and}
  \bibinfo{author}{\bibfnamefont{M.}~\bibnamefont{Lax}},
  \bibinfo{journal}{Phys. Rev. B} \textbf{\bibinfo{volume}{7}},
  \bibinfo{pages}{4491} (\bibinfo{year}{1972}).

\bibitem[{\citenamefont{Connor et~al.}(1979)\citenamefont{Connor, Hastie, and
  Taylor}}]{Connor}
\bibinfo{author}{\bibfnamefont{J.~W.} \bibnamefont{Connor}},
  \bibinfo{author}{\bibfnamefont{R.~J.} \bibnamefont{Hastie}},
  \bibnamefont{and} \bibinfo{author}{\bibfnamefont{J.}~\bibnamefont{Taylor}},
  \bibinfo{journal}{Proc. Roy. Soc. London Ser. A}
  \textbf{\bibinfo{volume}{365}}, \bibinfo{pages}{1} (\bibinfo{year}{1979}).

\bibitem[{\citenamefont{Chance et~al.}()\citenamefont{Chance, Dewar, Frieman,
  Glasser, Greene, Grimm, Jardin, Johnson, Manickam, Okabayashi
  et~al.}}]{Chaetal}
\bibinfo{author}{\bibfnamefont{M.~S.} \bibnamefont{Chance}},
  \bibinfo{author}{\bibfnamefont{R.~L.} \bibnamefont{Dewar}},
  \bibinfo{author}{\bibfnamefont{E.~A.} \bibnamefont{Frieman}},
  \bibinfo{author}{\bibfnamefont{A.~H.} \bibnamefont{Glasser}},
  \bibinfo{author}{\bibfnamefont{J.~M.} \bibnamefont{Greene}},
  \bibinfo{author}{\bibfnamefont{R.~C.} \bibnamefont{Grimm}},
  \bibinfo{author}{\bibfnamefont{S.~C.} \bibnamefont{Jardin}},
  \bibinfo{author}{\bibfnamefont{J.~L.} \bibnamefont{Johnson}},
  \bibinfo{author}{\bibfnamefont{J.}~\bibnamefont{Manickam}},
  \bibinfo{author}{\bibfnamefont{M.}~\bibnamefont{Okabayashi}},
 \bibnamefont{and}
\bibinfo{author}{\bibfnamefont{A.~M.~M.}~\bibnamefont{Todd}},
\bibinfo{note}{Proc. 7th International
  Conference on Plasma Physics and Controlled Nuclear Fusion Research,
  Innsbruck, Austria, 1978 (IAEA, Vienna, Austria, 1979) Vol. I, p. 677}.

\bibitem[{\citenamefont{Strauss}(1977)}]{Str}
\bibinfo{author}{\bibfnamefont{H.~R.} \bibnamefont{Strauss}},
  \bibinfo{journal}{Phys. Fluids} \textbf{\bibinfo{volume}{20}},
  \bibinfo{pages}{1354} (\bibinfo{year}{1977}).

\bibitem[{\citenamefont{Drake and Antosen~Jr.}(1984)}]{DraAnt}
\bibinfo{author}{\bibfnamefont{J.~F.} \bibnamefont{Drake}} \bibnamefont{and}
  \bibinfo{author}{\bibfnamefont{T.~M.} \bibnamefont{Antosen~Jr.}},
  \bibinfo{journal}{Phys. Fluids} \textbf{\bibinfo{volume}{27}},
  \bibinfo{pages}{898} (\bibinfo{year}{1984}).

\bibitem[{\citenamefont{Zaslavsky and Edelman}(2001)}]{ZasEde}
\bibinfo{author}{\bibfnamefont{G.~M.} \bibnamefont{Zaslavsky}}
  \bibnamefont{and} \bibinfo{author}{\bibfnamefont{M.}~\bibnamefont{Edelman}},
  \bibinfo{journal}{Chaos} \textbf{\bibinfo{volume}{11}}, \bibinfo{pages}{295}
  (\bibinfo{year}{2001}).

\bibitem[{\citenamefont{Garcia and Carreras}(2006)}]{GarCar06}
\bibinfo{author}{\bibfnamefont{L.}~\bibnamefont{Garcia}} \bibnamefont{and}
  \bibinfo{author}{\bibfnamefont{B.~A.} \bibnamefont{Carreras}},
  \bibinfo{journal}{Phys. Plasmas} \textbf{\bibinfo{volume}{13}},
  \bibinfo{pages}{022310} (\bibinfo{year}{2006}).

\bibitem[{\citenamefont{Samorodnitsky and Taqqu}(1994)}]{Taqqu}
\bibinfo{author}{\bibfnamefont{G.}~\bibnamefont{Samorodnitsky}}
  \bibnamefont{and} \bibinfo{author}{\bibfnamefont{M.~S.} \bibnamefont{Taqqu}},
  \emph{\bibinfo{title}{{Stable non-Gaussian processes}}}
  (\bibinfo{publisher}{Chapman \& Hall}, \bibinfo{address}{New York},
  \bibinfo{year}{1994}), \bibinfo{pages}{page 5}.

\end{thebibliography}
\end{document}